\documentclass[twocolumn, astrosymb]{aastex701}
\hypersetup{linkcolor=blue,citecolor=blue,filecolor=cyan,urlcolor=blue}

\submitjournal{ApJL}

\usepackage{amsmath,amstext}
\usepackage{amssymb}
\usepackage{hyperref}
\usepackage{graphicx}

\begin{document}

\title{Probing the large-scale magnetic field inside the Sun from three decades of observed surface magnetograms}

\author[orcid=0000-0001-7957-1727,gname=Soumyadeep, sname=Chatterjee]{Soumyadeep Chatterjee}
\affiliation{Department of Physics, Indian Institute of Technology Kanpur, Kanpur 208016, India.}
\email[show]{soumyade@iitk.ac.in}  

\author[orcid=0000-0001-5388-1233,gname=Gopal, sname=Hazra]{Gopal Hazra} 
\affiliation{Department of Physics, Indian Institute of Technology Kanpur, Kanpur 208016, India.}
\email[show]{hazra@iitk.ac.in}

\begin{abstract}

Space-weather and disturbances in the heliosphere are manifestations of the solar magnetic field, which is solely driven by the interior dynamo, and constraining the solar interior magnetic field and its oscillatory behavior is one of the major challenges in solar physics. Observationally, none of the techniques, including helioseismology, are able to provide an estimation of the interior magnetic field. We reconstruct, for the first time, the dynamics of the interior large-scale magnetic fields by assimilating observed line-of-sight photospheric magnetogram data from MDI/SOHO \& HMI/SDO along with helioseismic differential rotation data over three decades (1996-2025) into a 3D Babcock-Leighton dynamo model. The assimilation of observational magnetogram data allows us in realistic modelling of Babcock-Leighton mechanism as observed on the Sun without any simplified parameterization. As a result, our data-driven model successfully reproduces key observational features such as the surface butterfly diagram, accurate polar field evolution, and axial dipole moment. The reconstructed interior field dominated by toroidal component exhibits an equatorward migration and reproduces the realistic amplitude and modulation of cycles 23-25. We observe that the non-axisymmetric behaviour of the interior toroidal field becomes less prominent as we move deep towards the tachocline according to our model. A strong correlation between the simulated toroidal field and  sunspot number establishes our 3D magnetogram-driven model as a robust predictive model of the solar cycle.

\end{abstract}



\section{Introduction}\label{sec:introduction}
The solar cycle is nothing but a temporal variation of sunspots on the photosphere,  which emerge from deep-seated magnetic fields inside the Sun, especially from the dominant toroidal component of the magnetic field \citep[e.g.,][]{Parker55b,Cameron2018, Cameron2019}. Due to magnetic buoyancy, the interior toroidal field rises and produces the observed latitudinal and longitudinal distribution of sunspots along with the observed latitudinal dependence of tilt between sunspots \citep{DSilva93,Fan93}. Hence, understanding the interior magnetic field and its prediction is an important goal, as it dictates the surface manifestations of sunspots, solar cycles, and space-weather.

The most successful model to explain the solar cycle and the behaviour of the solar magnetic field is the mean-field flux transport dynamo model with Babcock-Leighton (BL) $\alpha$-effect \citep[e.g.,][]{Chou11, Char14, Hazra2023_review}. In this model, cyclic behaviour of the solar magnetic field arises from the continuous transformation between toroidal and poloidal magnetic components \citep{Parker55a, Radler_Dia_68}. Traditionally, BL models are based on an axisymmetric mean-field turbulence framework, but recently, there has been progress in developing a three-dimensional (3D) BL dynamo model with non-axisymmetric magnetic fields \citep[see for a review,][]{Hazra2021_review}. In the 3D model, the BL mechanism concentrated on the photosphere generates the poloidal field from sunspots that emerge from the interior toroidal field, and the toroidal field gets generated via $\Omega$-effect due to differential rotation from the existing poloidal field generated via BL mechanism, hence closing the dynamo loop. There are two main advantages of 3D models over existing 2D models while modelling the solar magnetic field. The BL mechanism and magnetic buoyancy, which are inherently 3D process, could be modelled more realistically and we can assimilate the observed photospheric data in great details \citep[e.g.,][]{HM18}.

The first development of a 3D kinematic dynamo model using a sophisticated buoyancy algorithm was performed by \citet{Yeates13}. They have calibrated their initial flux tube parameters with the observed distribution of BMRs and implemented them in simulating the global cycle. By coupling a 2D dynamo model and 2D surface flux transport model, \citet{Lemerle_2015ApJ} and \citet{Lemerle_2017ApJ} developed a hybrid model to simulate the global solar cycle as well as detailed surface flux transport. At the same time, \citet{Miesch_Dikpati_2014, Miesch_Teweldebirhan_2016} and \citet{HCM17} started developing the 3D dynamo model STABLE with spot deposition based on dynamo generated toroidal field but without explicit flux emergence, as they argued that the rising flux tube needs to be disconnected from the parent flux tube for sunspot stability and observed surface evolution \citep{Rempel2006, Whitbread2019}. This model can successfully reproduce the major features of the solar cycle, for example, 11-year periodicity, equatorward migration of sunspots, and polarity reversal. However, the realism of surface magnetic flux transport and polar field is far from the observations. This is because while modelling the BL mechanism, sunspots are modelled as two bipolar circular Gaussian patches and artificially suppressed in the higher latitudes. Not only in STABLE, but also in other 3D dynamo models, irrespective of different buoyancy algorithms, one thing to notice is that the BL mechanism on the surface has been modeled using empirical parameterization (sunspots as bipolar circular regions) guided by observations. It has already been shown that the bipolar approximation of sunspots without inherent asymmetry underestimates the polar field by 30-40\% in the Sun \citep{Iijima2019, yeates2020}.


Also, due to simple parameterization of BL process, in some cases, to reproduce realistic toroidal field that produces realistic solar cycle, a random scattered around the tilt angle have been introduced in the parametric BL mechanism \citep{KM17}, and sometime the simulated polar field from BL mechanism during minima of cycles - a well-established precursor to predict next cycle amplitude, has been increased or decreased based on observed polar field to predict next cycle amplitude \citep{CCJ07}. In addition, various quenching methods have been widely used in kinematic dynamo simulations to stop the unlimited growth of the magnetic field in the BL process, for example, tilt angle quenching \citep{KM17} and latitude quenching \citep{Jiang2020, Yeates2025}. These quenching mechanisms have also been incorporated in a rudimentary way \citep[e.g.,][]{KM17, Karak_2020ApJ}.

In this work, for the first time, we aim to model the realism of BL mechanism completely from observations by assimilating full photospheric magnetograms for three decades from 1996 to 2025 without any explicit parameterization of the process. This would probe the magnetic field inside the Sun very precisely, as by capturing the details of the BL mechanism accurately from surface magnetograms, we minimize the uncertainty arising from crude approximations of the mechanism and obtain a more accurate poloidal field. To be precise, this will particularly advance our understanding of the solar interior, as we do not have to synthetically model the BL process anymore by assuming the emergence of sunspots as two bipolar Gaussian patches and include tilt and/or latitude quenching parametrically. The inclusion of a full magnetogram that consists of morphological asymmetry of active regions, accurate latitudinal emergence, the real tilt angles, and weak diffuse fields intrinsically captures the realism of BL models with great accuracy and produces a realistic polar field. We discuss the details of the assimilation process in Section~\ref {sec:dataassimilation}. We present the results of surface flux transport and polar fields in Section~\ref {sec:polar_field}. 
Subsequently, the dynamo generated interior toroidal magnetic field from surface observations is presented in Section~\ref{sec:toroidal_dynamics}. How realistically the dynamo generated toroidal field resembles the cycle 23-25 and the predicting capability of our model are explained in Section~\ref{sec:cycle_prediction}. Finally, we present our conclusions in Section~\ref{sec:conclusion}.

\section{Magnetogram data, assimilation and the 3D Dynamo model} \label{sec:dataassimilation}

We have adopted the 3D kinematic dynamo model, STABLE, to simulate the magnetic field in the solar convection zone with axisymmetric flows and non-axisymmetric magnetic fields \citep{HCM17, HM18}. This model solves the induction equation: 

\begin{equation}
    \frac{\partial \mathbf{B}}{\partial t} = \nabla \times (\mathbf{V} \times \mathbf{B} - \eta \nabla \times \mathbf{B}) + \mathbf{S},\label{eq:induction_eq}
\end{equation}
in a three-dimensional (3D) rotating spherical shell with prescribed plasma velocity field ($\mathbf{V}$), turbulent diffusivity ($\eta$) (see Appendix.~\ref{sec:mcflow}\&\ref{sec:diffusivity} for the details) and a source term, $\mathbf{S}$, constructed from a parametrized sunspot deposition algorithm based on the deep toroidal field to mimic the BL mechanism. Numerically, STABLE is implemented by adapting the Anelastic Spherical Harmonic (ASH) code for solving only the induction equation. STABLE employs a pseudospectral method with a triangularly truncated spherical harmonic decomposition in the horizontal direction and a fourth-order finite-difference scheme in the radial direction. The time integration is performed using a mixed scheme (semi-implicit/explicit), whereas the smooth and regular behavior of the spherical harmonics obviates the need for any special treatment of singularities near the poles. Further numerical details can be found in \citet{Clune_1999, Brun_2004ApJ, Miesch_Dikpati_2014}.

Our simulation domain spans the solar convection zone from $0.69R_{\odot}$ to $R_{\odot}$ in radius, $0^\circ$ to $180^\circ$ in latitude, and $0^\circ$ to $360^\circ$ in longitude, where $R_{\odot}$ is the solar radius. The computational grid consists of 200, 256, and 512 points in the $r$, $\theta$, and $\phi$ directions, respectively. The mean magnetic field $\mathbf{B}$ in Equation~\ref{eq:induction_eq} is solved in terms of toroidal and poloidal potentials and we write $\mathbf{B}$ as
\begin{equation} \label{eq:decomp}
    \mathbf{B} = \nabla \times (A \hat{r}) + \nabla \times \nabla \times (C \hat{r}),
\end{equation}
where $A$ and $C$ represent the toroidal and poloidal potentials, respectively. We impose radial boundary conditions on the surface ($A = \partial C/\partial r = 0$), and perfectly conducting boundary conditions ($C = \partial A/\partial r = 0$) at the base of the convection zone, as described in \cite{Miesch_Teweldebirhan_2016}. 

Unlike previous models \citep[e.g.,][]{Miesch_Teweldebirhan_2016, HCM17} where $\mathbf{S}$ is calculated from dynamo generated toroidal field using `SpotMaker' algorithm, we construct the $\mathbf{S}$ using the observed line-of-sight (LOS) magnetogram data from the Michelson Doppler Imager (MDI) and the Helioseismic and Magnetic Imager (HMI) for our study. In this sense, our model is fully driven by the observational surface field. We assimilated LOS magnetic field measurements from MDI and HMI magnetograms from May 1996 to April 2025 with a 24-hour cadence for almost three decades. The observed MDI and HMI magnetogram data are available in different resolutions, and the magnetic information in the polar regions also has a lot of noise. We transform the LOS magnetogram data from the helioprojective into heliographic coordinates to get radial magnetic field $B_{r, obs}$ with appropriate center-to-limb correction using the  Magnetic Mapping and Processing (MagMAP) Python package \citep{MAGMAP, Caplan_2025} and down sampled them into our simulation grid. For the noise reduction, we have used the default weight function of MagMAP as follows:  

\begin{equation}
    W = \begin{cases}
        \mu^{\alpha} & \text{if} \ \mu > \mu_{lim} \\
        0, & \text{elsewhere}
    \end{cases}
\end{equation}

where, $\mu =\cos\Theta_l$, $\Theta_l$ is the center-to-limb angle, and $\mu_{lim}, \alpha$ are constants. We have chosen $\alpha=2$ for MDI and $\alpha=1$ for HMI data. The value of $\mu_{lim}$ is 0.1 in both cases. While MagMAP was originally developed for HMI data, we have extended its functionality to support MDI as well. The reduced radial magnetic field map used in the STABLE model is given below:
\begin{equation}
    B_{r, reduced}(r = R_\odot,\theta, \phi,t) = B_{r, obs}(\theta, \phi,t) W(\theta, \phi,t).\label{eq:ass_form} 
\end{equation} 

\begin{figure*}[!htbp]
   \centering
   \includegraphics[width=1.0\linewidth]{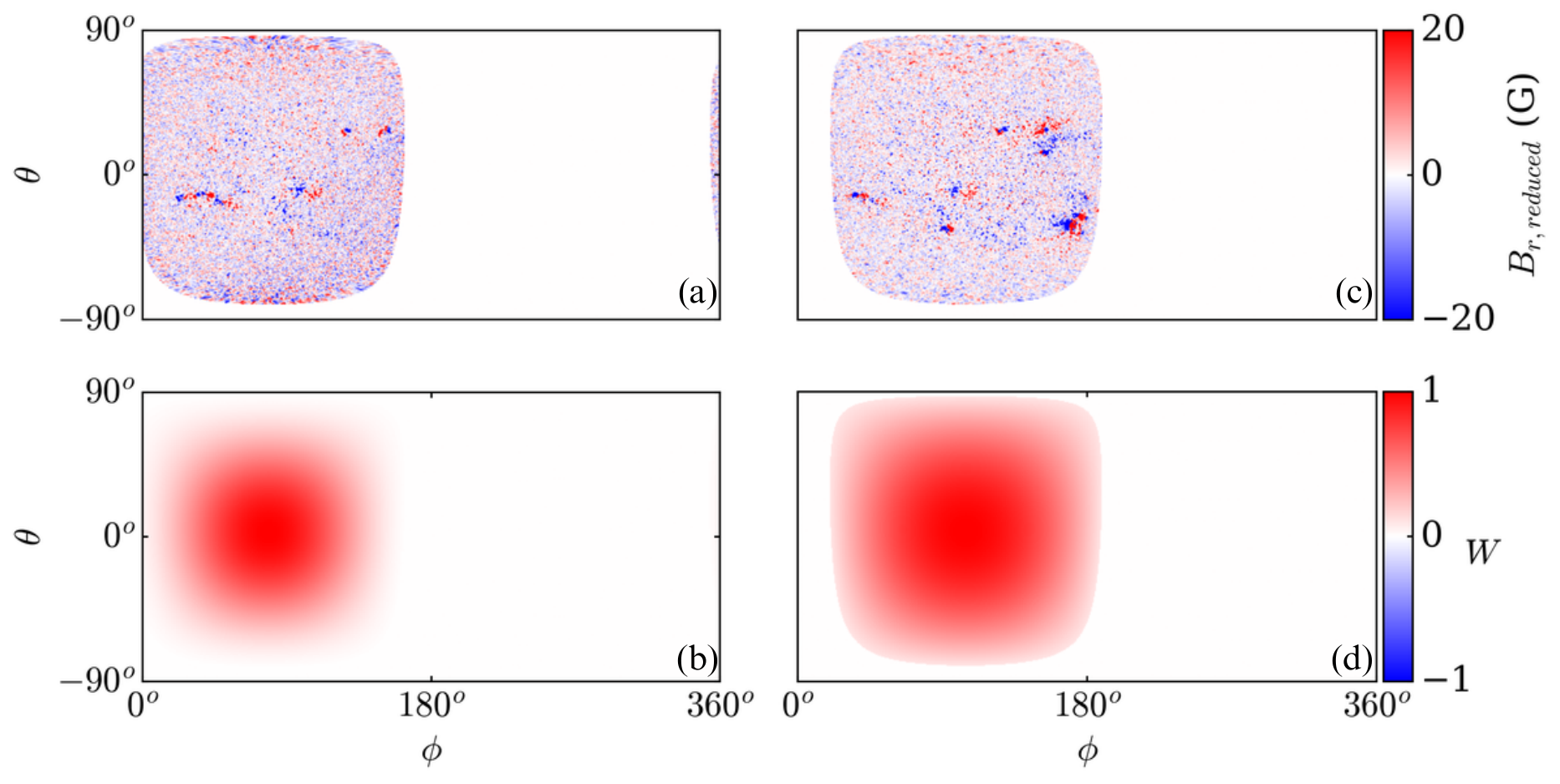}
   \caption{(a) Assimilated \(B_{r,reduced}\) in the STABLE model corresponding to the MDI magnetogram observed on December 30, 2010, after processing with the MagMAP package and applying the weight factor shown in (b). (c) Assimilated \(B_{r,reduced}\) corresponding to the HMI magnetogram observed on January 1, 2011, processed similarly using MagMAP with the weight factor shown in the panel (d).}
   \label{Fig:assimilated_magnetogram_mdi_hmi}%
\end{figure*}

Figure~\ref{Fig:assimilated_magnetogram_mdi_hmi}(a,b) and (c,d) show examples of the reduced radial magnetic field ($B_{r, reduced}$) from MDI and HMI, respectively, along with their corresponding weight functions. An important reason for choosing the value of $W$ in this way is to minimize noise. For MDI, the noise amplitude is highest in the southwest part of the disk \citep{Liu_2012}, whereas HMI magnetograms exhibit a uniform noise level of approximately 5 G \citep{Couvidat_2016, Boucheron_2023}. To balance the flux in HMI data with MDI, we multiply Equation~\ref{eq:ass_form} by 1.4 for HMI data, following \citet{Liu_2012}. We should also mention that the $B_\theta$ component from observation has not been considered in our assimilation, as $B_\theta$ is much less compared to the $B_r$ at the surface \citep[e.g.,][]{Vidotto_2016}.

Surface magnetic fields, especially sunspots, are expected to be anchored below the surface up to a certain depth, but they need to be disconnected from the parent flux tube at the tachocline \citep{Kosovichev2006, Rempel2006, Whitbread2019} to explain surface magnetic field evolution and stability. While assimilating the magnetogram
data, we have extrapolated the surface fields up
to a depth of 0.98$R_\odot$ ($\simeq $14 Mm) from the surface using a simple radial extrapolation function.
One of the motivations of this extrapolation is to
keep the BL mechanism confined near the surface \citep{Bab61, Lei69, DC99, Chatterjee_2004}. Also, we chose the extrapolation depth up to $0.98R{_\odot}$ as earlier studies \citep{Rempel2006, Whitbread2019, Strecker21} found that sunspots should be disconnected at around $\sim$ 10 Mm distance from the surface for maintaining stability
and the observed surface magnetic field evolution. However, the depth of this subsurface connectivity is not uniquely constrained by both numerical modelling and helioseismic observations \citep{Moradi2010SoPh, Kosovichev_2012SoPh, Fan_2021LRSP}. Previous studies by \citet{Rempel2006, Whitbread2019, Strecker21} suggested that sunspots become effectively disconnected at depths of the order $\sim$ 10 Mm below the surface, whereas radiative magnetohydrodynamics simulation spanning the full convection zone by \citet{Hotta2020MNRAS} showed that sunspots can retain much deeper connectivity within the convection zone. The connectivity of the sunspots in the convection zone remains an active field of research. For a detailed study of how the surface evolution of the magnetic field depends on the subsurface connectivity of the sunspots, we refer to \citet{Whitbread2019}.

For the subsurface structure of surface fields, the reduced magnetogram data, $B_{r,reduced}$ (as given in Equation~\ref{eq:ass_form}) is extrapolated below using the following empirical relation:
\begin{equation}
    B_{r, extrapol} = \frac{B_{r, reduced}}{2} \frac{\tanh(Ar - B)}{\tanh(A - B)} + \frac{B_{r,reduced}}{2}.\label{eq:potential_extrapolation}
\end{equation}
The hyperbolic tangent function simplifies the numerical implementation in real space while maintaining an extrapolation profile similar to that used by earlier studies \citep{Miesch_Dikpati_2014, Miesch_Teweldebirhan_2016, HM18}. It provides a smooth transition of the non-zero surface magnetic field to zero at $r = 0.98 R_{\odot}$, with constants $A = 95$ and $B = 90.25$. Equation~\ref {eq:potential_extrapolation} takes care of the extrapolation of the diffuse field as well. 

\begin{figure}[!htbp]
	\centering
	\includegraphics[width=0.45\textwidth]{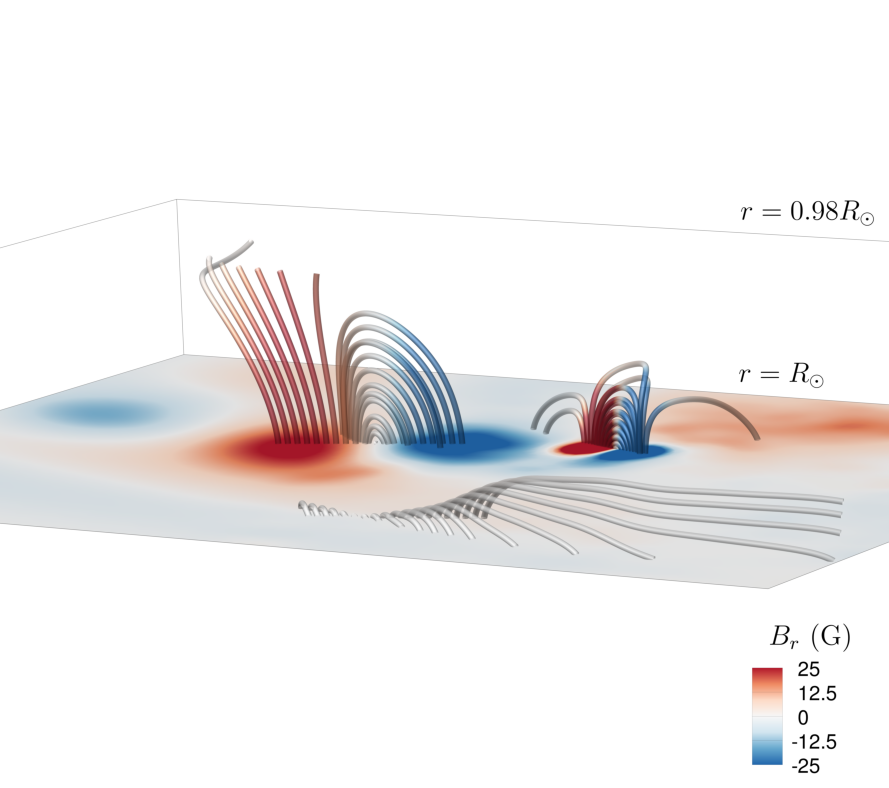} 
	\caption{Subsurface extrapolation of active regions inside the convection zone. Red and blue colors show the positive and negative polarities of active regions. Active regions are extrapolated deeper in the convection zone up to $0.98R_\odot$ based on their sizes. The bigger active regions go up to a maximum $0.98R_\odot$, but other active regions get extrapolated in between the surface and $0.98R_\odot$ based on their area. The solar surface is shown using the filled contour and extrapolation below the convection zone is shown vertically upward using streamlines till $0.98 R_\odot$.}
	\label{fig:extrapolation} 
\end{figure}

Once we obtained the magnetogram data B$_{r, extrapol}$ with extrapolation up to 0.98R$_\odot$, we transform it to a source term $S_C$ for poloidal potential $C$. In our framework, we only include the source for the poloidal potential $C$. Since we assimilate solely the radial magnetic field, the source term for the toroidal potential $A$ becomes identically zero. We calculate $S_C$ as given below:
 
\begin{equation}
    \tilde{S}_C = \frac{r^2}{l(l+1)} \tilde{B}_{r,extrapol}, \label{eq:poloidal_pot_compute}
\end{equation}
where $\tilde{S}_C$ and $\tilde{B}_{r,extrapol}$ are obtained via spherical harmonic transforms in latitude and longitude from $S_C$ and $B_{r,extrapol}$, respectively. See Appendix~\ref{sec:source term} for further details on the derivation of Equation~\ref{eq:poloidal_pot_compute}. This completes our data assimilation process.
In this formulation, we assume that the subsurface field up to the extrapolated region is purely radial. This is a simplified assumption, but in our considered extrapolation depth, the field lines are expected not to diverge from radial structure \citep{Strecker21}. 
Figure~\ref{fig:extrapolation} shows a typical extrapolation near sunspot regions. Note that the penetration depth of the extrapolated surface field in the extrapolated region up to 0.98R$_\odot$ has a dependency on the spherical harmonic number {\em l}, which makes sure that the surface field with a strong sunspot magnetic field penetrates deeper, whereas the diffuse field does not penetrate all the way up to 0.98R$_\odot$.

\begin{figure*}[!htbp]
	\centering
	\includegraphics[width=1.0\textwidth]{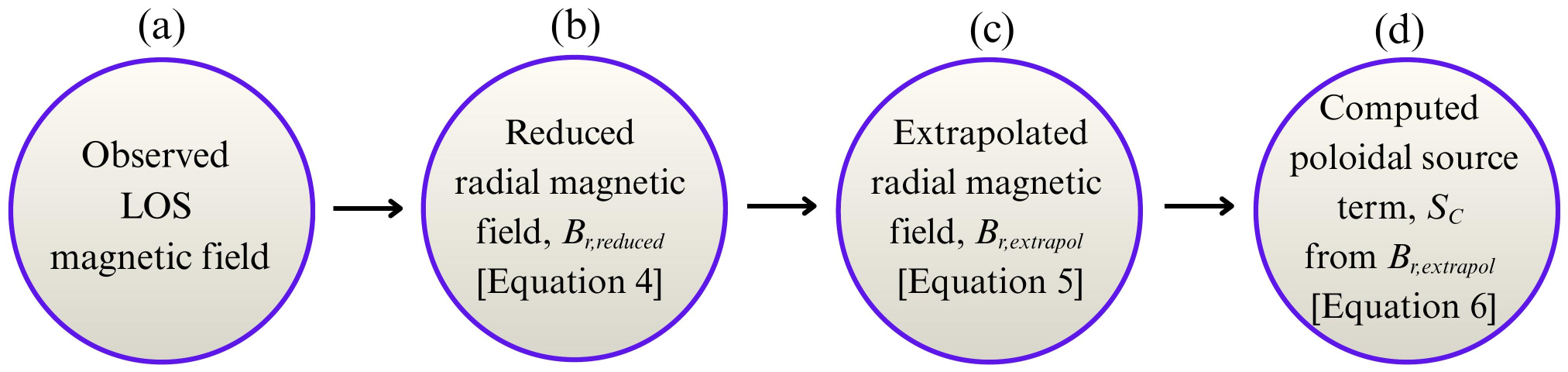} 
	\caption{Schematic representation of the data assimilation procedure implemented in our model.}
	\label{fig:schematic_assim} 
\end{figure*}

In summary, we show our assimilation method step by step in Figure~\ref {fig:schematic_assim}. In step (a), we obtain the LOS magnetogram data from the observation first. Then we reduce the data to be used in our simulation grid by converting them to heliographic coordinates, correcting for noise, and downsampling them into our simulation grid using MagMAP (step (b)). After we obtain reduced data, the next step (c) is to build the subsurface structure of the surface field, which is computed using a simple radial extrapolation technique following Equation~\ref{eq:potential_extrapolation}. Once we have radial magnetic field data in the extrapolation region, we use them to calculate the poloidal source term ($S_C$) using Equation~\ref{eq:poloidal_pot_compute} (step (d)). This poloidal source term is then finally included in the induction equation to study the global evolution of all components (B$_r$, B$_\theta$ and B$_\phi$) of magnetic field.

\begin{figure}
	\centering
	\includegraphics[width=0.5\textwidth]{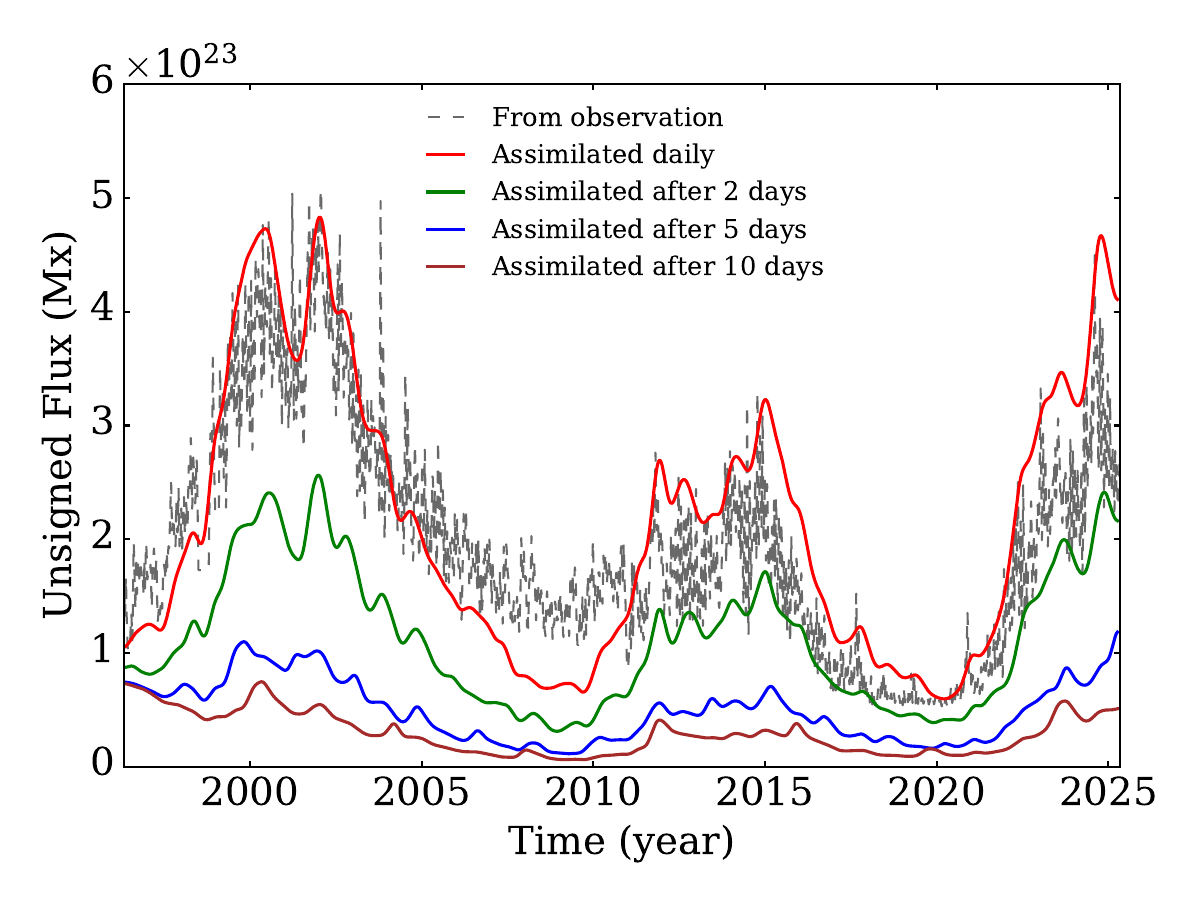} 
	\caption{Time evolution of the unsigned radial magnetic flux from daily magnetic maps observed by MDI and HMI (gray), compared with simulations using assimilation cadences of 1 day (red), 2 days (green), 5 days (blue), and 10 days (brown).}
	\label{fig:data_assim} 
\end{figure}

After fixing the active regions' depth, the only parameter left in assimilating data was to find out the assimilation time interval. By optimizing the simulated $B_r$ from the assimilated $B_{r, obs}$ of the present time step, against the observed magnetic field $B_{r, obs}$ at the next time step as a function of time, we find that daily assimilation of magnetogram data is only able to conserve the surface flux without much diffusion.
Hence, we allow the relaxation time of $B_r$ for a day for surface flux to be conserved. Figure~\ref{fig:data_assim}(b) shows the time evolution of the unsigned magnetic flux, where the red, green, blue, brown, and gray lines represent simulations with 1-day, 2-day, 5-day, and 10-day assimilation cadences and observations, respectively. The close agreement between the daily assimilation run and the observations confirms that the assimilation conserves magnetic flux accurately. We have also incorporated real-time differential rotation that includes near-surface shear layers from helioseismic observations based on 72-day time series from MDI and HMI (data obtained via private communication with Prof. H M Antia \citep{Antia_2022}, and a snapshot on 1st May 1996 is shown in Figure~\ref{Fig:assimilated_dfr_mdi}) to realistically model internal dynamo.  The meridional flow and turbulent diffusion are taken from helioseismic measurements using analytic formulas from \cite{HKC14, HCM17}. Also, it is worth mentioning that assimilation of surface $B_{r, obs}$ ensures that the sunspot field automatically gets inserted in the dynamo model, and we do not need explicit sunspot emergence from the dynamo model using the self-consistent buoyancy algorithm. In our case, the observational sunspot fields act as a source function of poloidal potential and generate a toroidal field due to the induction effect.

\section{Surface flux transport and polar field: 1996-2025}\label{sec:polar_field}

\begin{figure*} 
	\centering
	\includegraphics[width=1.0\textwidth]{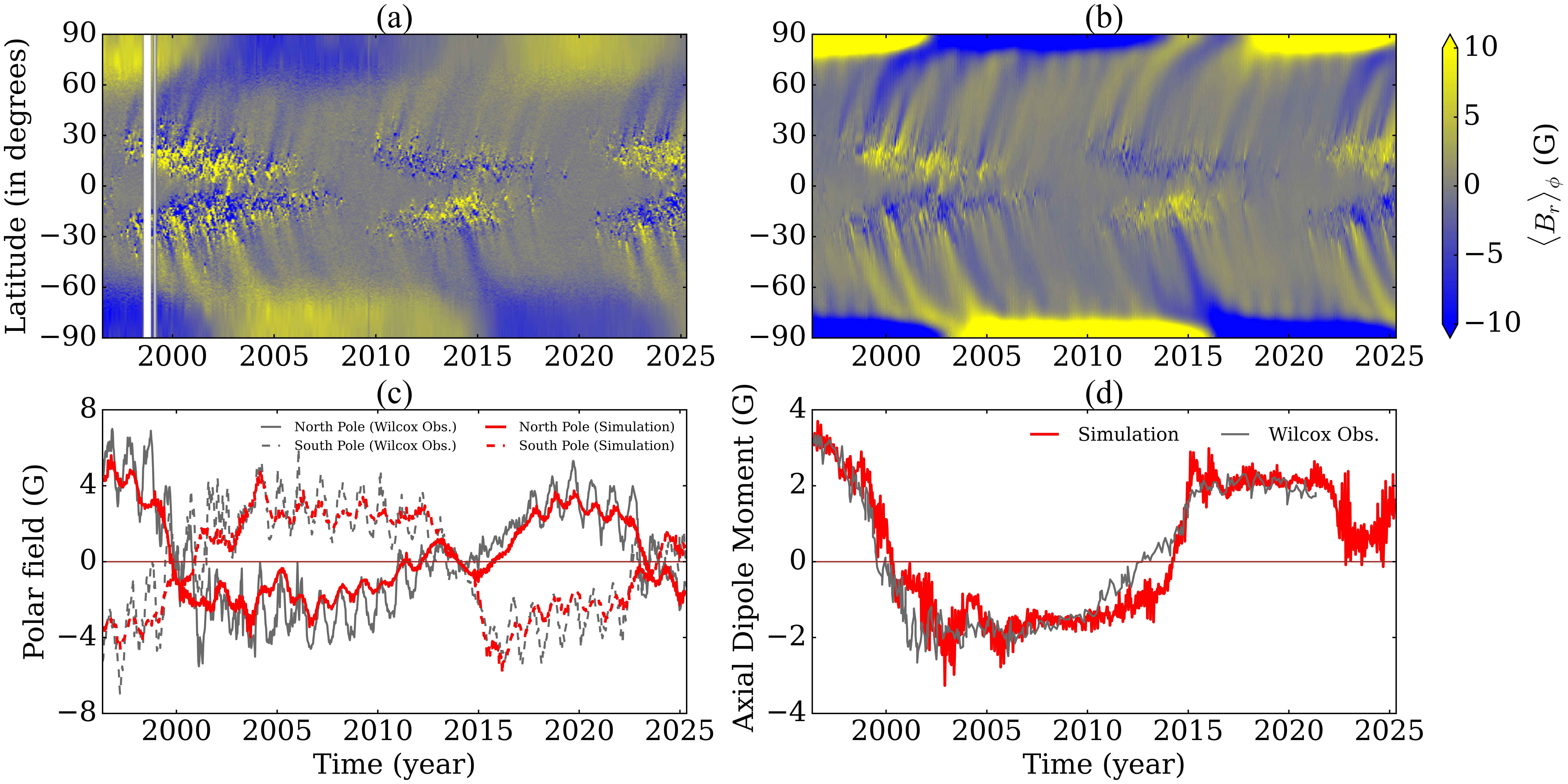} 
	\caption{Evolution of the longitude-averaged radial magnetic field at the Solar surface - commonly known as the butterfly diagram from (a) Observational data from 1996 to 2025, derived from MDI and HMI magnetograms, and (b) Simulated data from 1996 to 2025, obtained from our simulation. (c) Time evolution of the polar magnetic fields from our simulation (red) compared with observations from the Wilcox Solar Observatory (gray). Solid lines represent the northern polar field; dashed lines represent the southern polar field. (d) Same as (c) but for the axial dipole moment.}
	\label{fig:surface_bfly}
\end{figure*}

By successfully assimilating 30 years of full-disk daily magnetogram data into the 3D dynamo model, we are able to simulate realistic surface evolution of the magnetic field, polar field, and axial dipole moment as observed. In Figure~\ref{fig:surface_bfly}(a,b), we show the surface flux transport from 1996-2025 from the observation and from our assimilated model, respectively. The simulated butterfly diagram shows good agreement with the observations in terms of the latitudinal migration and polarity patterns. However, the simulated diagram appears smoother and less granular than the observed one. This difference is primarily due to the low spatial resolution used in our simulation. The magnetogram resolution was downsampled to $256 \times 512$. A higher spatial resolution in our simulation produces a more granular structure similar to observations, but achieving this for all considered cycles would require significantly higher computational cost. It is also worth noting that during the ``SOHO summer vacation" (shown by the white strip in Figure~\ref{fig:surface_bfly}(a)) period, we continued assimilating the last available magnetograms with an exponential decay function until the new data was recovered. This step ensured continuity and prevented artificial offsets in both polar and toroidal field evolution.

The polar field and axial dipole moment from our data assimilated model are shown in Figure~\ref{fig:surface_bfly}(c) and (d), respectively. The polar fields at both poles are calculated by integrating $B_r$ from $\pm 55^{\circ}$ to the pole. The observational values of these quantities from Wilcox Solar Observatory data are also overplotted using gray lines in the respective plots. Our simulated polar fields in both hemispheres match perfectly with the observations. The simulated axial dipole moment, which is a robust entity to quantify the global magnetic field of the Sun, has an excellent match with the observation. It is crucial to note that existing 2D Surface Flux Transport (SFT) Models that include sunspot area data to model active regions struggle to simulate the polar field close to the observations, including cycle-to-cycle variations, largely because of uncertain measurements in tilt angles of sunspot groups \citep{Jiang_2014,Jiang15}. Once the polar field is successfully simulated, the peak amplitude of the polar field is used to predict the next cycle amplitude via a well-established correlation \citep{CCJ07, Hazra_2019_pred}. Although via correlation one can have an estimation of the peak amplitude of the next cycle, one should self-consistently simulate the next cycle by considering the action of differential rotation, meridional flow and turbulent diffusion together on the BL mechanism to compute the realistic polar field and hence detailed timing and estimation of its shape and peak value, which is successfully executed in our model.

\section{Estimation of toroidal magnetic field inside convection zone}\label{sec:toroidal_dynamics}

\begin{figure*} 
	\centering
	\includegraphics[width=1.0\textwidth]{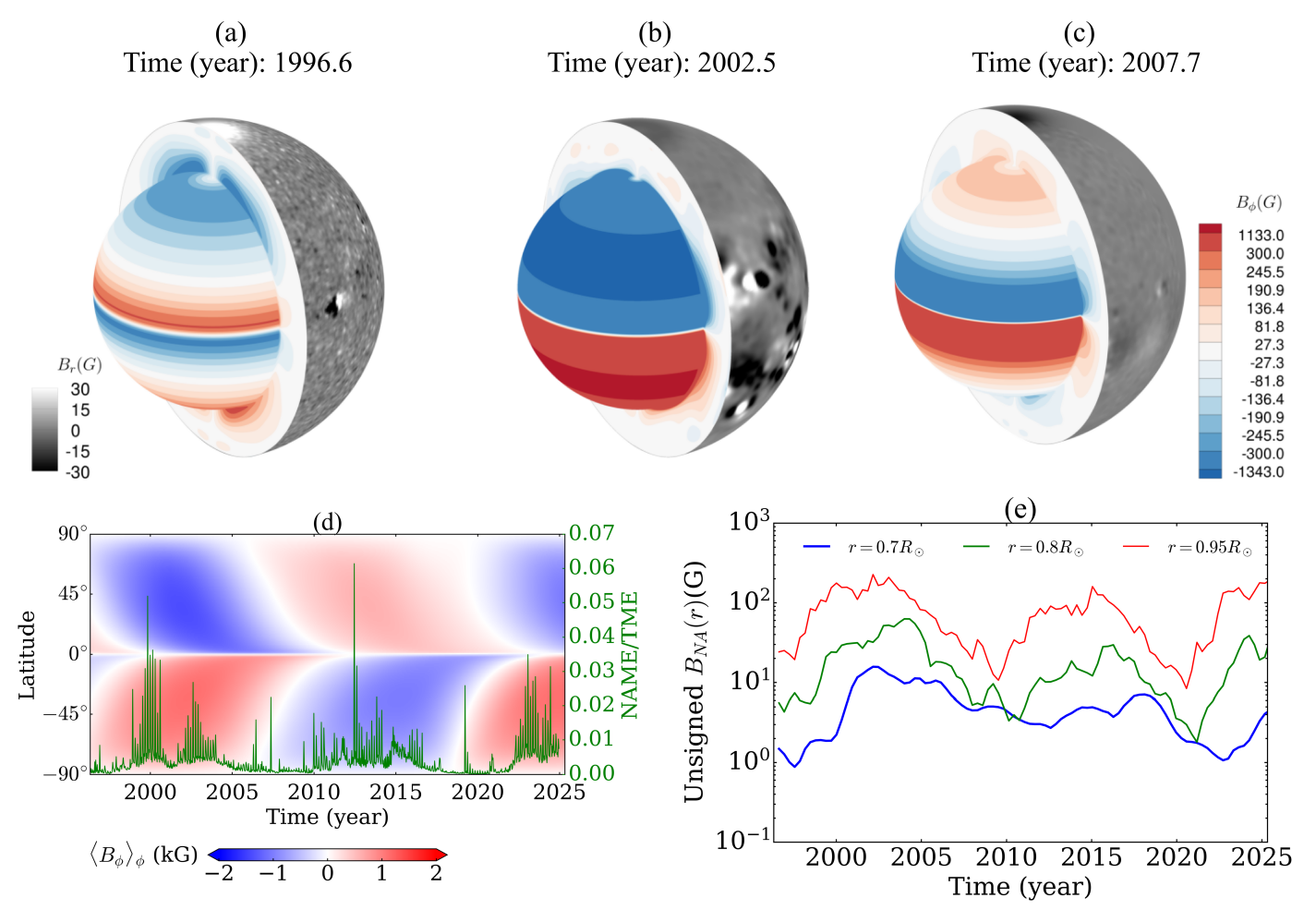} 
	\caption{Top panel, (a), (b), (c) show the 3D evolution of the toroidal field during cycle 23 at three different times. The filled contours show the toroidal field in the tachocline (r = 0.7R$_\odot$) in the spherical surface as well as in the meridional plane of the convection zone. Black and white filled contours on the surface of the Sun show the surface radial field.  (a), (b), (c) show the behaviour of toroidal and radial field near minimum (1996.6 year), at maximum (2002.5 year), and near the next minimum (2007.7 year) respectively encompassing the whole cycle 23. (d) Time-latitude plot of azimuthal averaged toroidal field ($\langle B_\phi \rangle_\phi$) that shows the equatorward migration.  The green line shows the evolution of the non-axisymmetric toroidal magnetic energy (NAME) relative to the axisymmetric toroidal magnetic energy (TME). (e) Here we show the contribution of non-axisymmetric toroidal field ($B_{NA}$) at different depths of the convection zone. The red, green, and blue lines show the magnitude of the non-axisymmetric magnetic field near the surface (0.95R$_\odot$), mid-convection zone (0.8R$_\odot$), and at the bottom of the convection zone (0.7$R_\odot$). }
	\label{fig:tor_full} 
\end{figure*}

We calculate all three components of the magnetic field inside the convection zone, generated by the combined effect of differential rotation and turbulent diffusion on the surface assimilated poloidal field from our model. In Figure~\ref{fig:tor_full}(a),(b) and (c), we show the 3D dynamics of the toroidal field inside the convection zone, and the radial field on the surface at different stages of cycle 23. Figure~\ref{fig:tor_full}(a) shows the toroidal field just after solar minimum of cycle 23, where we can see bands of toroidal field just starting to develop from the pole region in the higher latitudes, whereas the opposite polarity toroidal field from the previous cycle is diminishing with a final reminiscence in the low latitudes. In Figure~\ref{fig:tor_full}(b), we show the toroidal field at the maxima of cycle 23. We can see here that the toroidal field that started to develop at the higher latitude during minima has moved towards the equator, and both hemispheres have now opposite polarity. As time progresses, the new polarity of the toroidal field from the next cycle starts to develop from the polar region due to differential rotation, as shown in Figure~\ref{fig:tor_full}(c). During the progress of the toroidal field with time, the high latitude toroidal bands move to the equator, giving rise to the equatorward migration of the toroidal field. This becomes clear from Figure~\ref{fig:tor_full}(d), where we have shown the time-latitude variation of the azimuthally averaged toroidal field over the tachocline.

The toroidal field is more confined near the tachocline (as shown by the meridional plane of 3D figures in the Figure~\ref{fig:tor_full}(a),(b) and (c) due to both radial shear and latitudinal shear of differential rotation, and decreases towards the surface. Near the surface, the poloidal field dominates. In Figure~\ref{fig:tor_full}(d), we show the contribution of the total non-axisymmetric toroidal energy to total dynamo-generated toroidal magnetic energy in the green line over the butterfly diagram. We find that during solar maxima, the non-axisymmetric component becomes strong and contributes almost 7\% of the total toroidal magnetic energy. The main contribution to this non-axisymmetric component directly comes from the assimilated surface radial magnetic field distribution.  
This is a first estimate of the total non-axisymmetric component of the solar magnetic field inside the Sun from surface observations. We have also plotted the non-axisymmetric contribution of toroidal field from the assimilated magnetogram at different depths in Figure~\ref{fig:tor_full}(e). We find that it decreases as we go towards the tachocline (See Appendix~\ref{sec:nas_comp} for latitudinal and longitudinal variations at different times). This is not surprising as radial shear due to differential rotation being stronger at the tachocline twists the poloidal field in the azimuthal direction, making it axisymmetric. However, note that this is an estimate from a purely kinematic dynamo model that does not incorporate any explicit non-axisymmetric mechanism (e.g., tachocline instability) and convection. Hence, this might not be a true estimate of the non-axisymmetric component of the total toroidal field. Further study needs to be carried out in that direction, which will  include tachcocline instability. 

\begin{figure*} 
	\centering
	\includegraphics[width=1.0\textwidth]{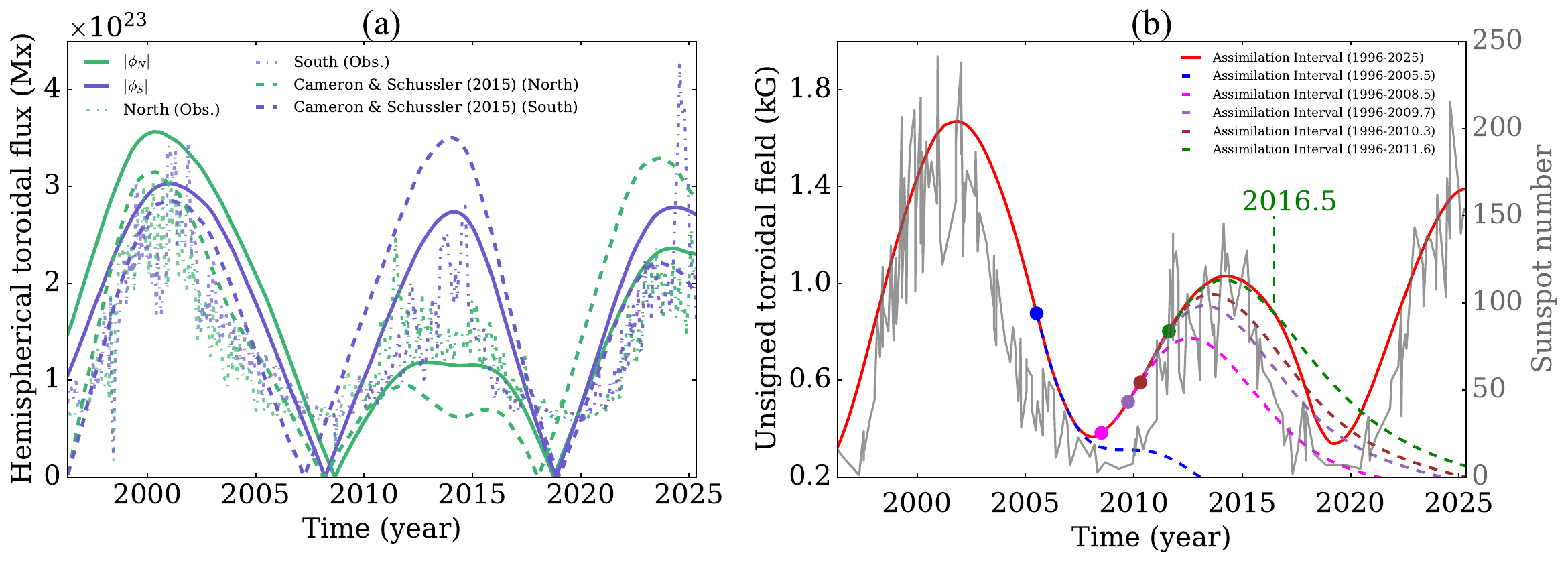} 
	\caption{(a) Time evolution of the toroidal flux inside the convection zone in both hemispheres from our simulation using solid lines. Green and blue colors show the toroidal fluxes of the northern ($\phi_N$) and southern hemisphere ($\phi_S$), respectively.  We have also overplotted the toroidal flux using dashed lines from the prescription of \cite{cameron2015} that combines surface magnetic field and differential rotation. The unsigned surface magnetic flux from MDI and HMI is also overplotted using dash-dotted lines to compare our results with observations. (b) Unsigned toroidal field from our simulation, as a proxy for the solar cycle, using a solid red line. The monthly averaged sunspot numbers are also overplotted with a solid gray line. The blue, magenta, cyan, purple, brown, and green dashed lines show the evolutions of the simulated toroidal fields when the assimilation is stopped at the following times: 2005.5 year, 2008.5 year, 2009.7 year, 2010.3 year, and 2011.6 year.}
	\label{fig:tor_flux} 
\end{figure*}

We estimated the total toroidal flux generated inside the convection zone with time as shown in Figure~\ref{fig:tor_flux}(a). The toroidal flux in the southern and northern hemispheres is shown in green and blue solid lines. The estimated toroidal flux varies with time as well as with hemisphere. This is a more reliable estimate of toroidal flux inside the convection zone from observations to date, compared to previous estimates \citep[e.g.,][]{cameron2015}. In order to validate our result with direct observation, we have also plotted the observed unsigned radial magnetic flux from the synoptic magnetogram for each hemisphere by a dotted line on the same figure. As the subsurface toroidal flux emanates in terms of sunspots, toroidal flux inside the convection zone should be comparable with the directly observed radial flux \citep{Cameron2018}. We indeed see that our simulated toroidal flux matches very well with observations. Earlier, there was an effort to estimate the toroidal field inside the convection zone from the surface field and surface differential rotation using Stoke's theorem \citep{cameron2015}. We compare our toroidal flux with their prescribed estimation, which is shown in green and blue dashed lines for the northern and southern hemispheres. Although qualitatively, their estimated toroidal flux matches ours, we notice some deviation. This is due to cross-equatorial diffusion of flux near the surface, which was not accounted for in their calculation \citep{cameron2015}. The cross-equatorial diffusion (see Appendix~\ref {sec:cross-equtorial} for their estimation) near the solar maxima is stronger compared to minima (see Figure~\ref{Fig:cross_equatorial}(b)), resulting in a stronger deviation of toroidal flux from their model to our estimation. 

Although our model provides first time an estimation of the toroidal field, we should note that this is an estimate of the mean toroidal flux inside the convection zone. Therefore, intermittency in the toroidal field could not be captured in our model \citep{Chou2003}. Also, one might think about how sensitive these values are compared to the consideration of different turbulent parameters, for example, turbulent diffusion and meridional flows. We have considered a two-step diffusivity profile by fixing surface diffusivity ($\eta_t$), diffusivity in the convection zone ($\eta_{mid}$) and the diffusion in the tachocline ($\eta_c$) as shown in the first row of Table~\ref{tab:eta} (see details in Appendix~\ref{sec:diffusivity}) for our calculation. We ran a few test cases by changing turbulent diffusivity inside the convection zone as shown in Table~\ref{tab:eta}, and we find that the peak amplitude of the signed toroidal field varies but not very significantly. A change in the meridional flow does not significantly change the magnitude of toroidal flux either, but it results in a slight shift in toroidal flux reversal and hence cycle period as expected from flux-transport dynamo with BL-$\alpha$ \citep{DC99}. The effect of diffusivity and meridional circulation on our results follows a similar behaviour as previous 2.5D flux transport dynamo models \citep[for details see,][]{yeates2008}.

\begin{table}[]
    \centering
    \begin{tabular}{c|c|c|c|c}
    \hline
    Two-step  & $\eta_c$ & $\eta_{mid}$ & $\eta_{top}$ & Maximum $B_\phi$ \\ 
    diffusivity & [cm$^2$s$^{-1}$]  & [cm$^2$s$^{-1}$]  & [cm$^2$s$^{-1}$] & (kG)\\
    \hline
    Standard & $2\times 10^{10}$ & $5 \times 10^{11}$ & $3 \times 10^{12}$ & 1.4  \\  
    \hline
    Test-1 &  $10^9$ & $5 \times 10^{11}$ & $3 \times 10^{12}$ & 3.0 \\ 
    \hline
    Test-2 &  $10^9$ & $5 \times 10^{10}$ & $3 \times 10^{12}$ & 6.5 \\ 
    \hline
    Test-3 &  $5 \times 10^8$ & $5 \times 10^{10}$ & $3 \times 10^{12}$ & 6.5 \\
    \hline
    Test-4 & 10$^9$ &  $1 \times 10^{11}$ & $1 \times 10^{12}$ & 5.5 \\
    \hline
    \end{tabular}
 \caption{Dependency of dynamo generated toroidal field on different turbulent diffusivity used in our model.\label{tab:eta}}
\end{table}

\section{Reproducing Solar cycle 23-25: Predictive capability of our model}\label{sec:cycle_prediction}
A realistic estimation of toroidal flux would be able to reproduce the quantitative features of the solar cycle, and indeed, our toroidal flux does so. In Figure~\ref{fig:tor_flux}(b), we show the toroidal field using a red solid line at the tachocline, which is used as a proxy for the sunspot cycle, overplotted with the observed sunspot numbers (gray line). We see that our observationally estimated toroidal flux accurately captures the peak amplitude of all three cycles. However, as we kept the meridional circulation constant over all three cycles, there are differences in the reversal of cycles in our model by a few months. This can be easily improved by implementing a variation of meridional circulation (\cite{Hazra_2019_pred}).

We also run a few simulations to investigate how early we can predict the cycle amplitude. We have stopped assimilating our magnetogram a few years before the cycle maxima to check how early we can predict the peak and timing of the amplitude. This is not an attempt to directly predict individual sunspots or polar fields, but rather to demonstrate that the internally generated toroidal field remains physically consistent with the cycle amplitude once assimilation is stopped. Unlike existing SFT based and 2D dynamo-based prediction schemes \citep{Labonville_2019, Jiang_2023, Bhowmik_2023, Jouve_2025}, our model is qualitatively different in both design and diagnostic capability, as we assimilate the full surface magnetogram and track the subsequent evolution of the complete 3D magnetic field.

We present five examples in Figure~\ref{fig:tor_flux}(b), where assimilation stopped at 2005.5 year (decay phase), 2008.5 year (during minima), 2009.7 year (early rising phase), 2010.3 year (rising phase) and 2011.6 (late rising phase), which are marked using filled blue, magenta, purple, brown and dark green circles respectively. The corresponding toroidal field evolution after assimilation was stopped at five different epochs are overplotted using blue, magenta, purple, brown and dark green dashed lines. As it is evident from the plot that the simulation with assimilation terminated in the middle of 2011 (shown using dashed dark green line) successfully predicts the peak amplitude of Cycle 24 and agrees closely with the red solid line for nearly five years after assimilation is stopped. Also, the run with assimilation ending in the middle of 2005 correctly predicts the timing of the Cycle 23 minimum. All five cases that we present here could predict the behaviour of the toroidal field approximately between 3-4 years in advance after the assimilation was terminated. Hence, to predict the solar maxima, we need to stop assimilating data as early as 3 years before the maxima for a reasonable prediction. As toroidal flux is calculated from the polar field that gets transported to the tachocline by the turbulent diffusivity and meridional circulation, we find that a few years (as early as 3 years) before maxima, is optimum with our chosen transport parameters to predict the peak of all cycles properly by only assimilating magnetogram data. Note that our aim here is not to present a quantitative prediction scenario in this paper for any cycle; rather, we aim to establish the fact that our model has the capability to predict future solar cycles by only assimilating surface magnetogram data.


\section{Conclusion}\label{sec:conclusion}

In summary, for the first time, we have calculated the Babcock-Leighton source term from observed magnetogram data assimilation in a 3D dynamo model without an uncertain buoyancy mechanism and parameterization of sunspot deposition, to reconstruct the toroidal field inside the convection zone of the Sun. By assimilating observed surface magnetograms from MDI and HMI and incorporating helioseismic measurements of differential rotation, we performed realistic simulations of the solar magnetic cycle spanning solar cycles 23 to 25. We successfully reproduce the realistic surface flux transport, observed polar field, including its cycle-to-cycle variation. The high-latitude reconstruction of surface flux transport is tremendously important because it shapes the global corona and builds the seed field for the next cycle.

The reconstructed toroidal field shows equatorward migration and produces realistic cycles with their amplitude modulation. We observationally constrained the non-axisymmetric component of the solar magnetic field, which shows a cyclic dependence and becomes stronger near the surface and during solar maxima. This is a first observationally guided reconstruction of the toroidal field inside the convection zone. We also showed that data assimilated dynamo model could be used to predict the cycle peak amplitude and time, as much as 3-4 years before the arrival of solar maxima. By assimilating observational data in a 3D dynamo model, our methodology paves the way for future temporal as well as latitudinal prediction of active region emergence on the surface and details of the solar cycle.  

\begin{acknowledgments}
The authors would like to thank two anonymous referees for their constructive comments that helped a lot to improve our manuscript. The computations were performed on the Param Sanganak supercomputer at IIT Kanpur. Access to Param Sanganak is provided by the National Supercomputing Mission and Indian Institute of Technology Kanpur. Some of the computations were also performed in the Pegasus cluster, IUCAA Pune. The authors thank Prof. H M Antia for kindly providing the differential rotation data for the last three decades. The HMI data used are courtesy of the NASA/SDO and the HMI science team. We used MDI data onboard SOHO, which is a project of international cooperation between ESA and NASA. We thank Arnab Rai Choudhuri, Vaibhav Pant, Mark Miesch, Adam Finley, Dibyendu Nandi, Bibhuti Jha, Sahel Dey, Tanay Srivastava, and Subhadip Pal for useful discussions and comments. We also thank Antoine Strugarek for kindly providing some plotting routines. The authors thank funding from the IIT Kanpur initiation grant IITK/PHY/2022386.
\end{acknowledgments}

\facilities{SOHO(MDI), SDO(HMI)}

\software{MagMAP \citep{MAGMAP}}


\appendix

\section{Details of source term}\label{sec:source term}

Here we describe the construction of the source term, $\mathbf{S}$. Using Equation~\ref{eq:decomp}, we rewrite the Equation~\ref{eq:induction_eq} in terms of $A$ and $C$ as
\begin{eqnarray}
    \frac{\partial C}{\partial t} &=& -r^2 \mathcal{L}^{-2}(\hat{r} \cdot[\nabla \times (\mathbf{V} \times \mathbf{B} - \eta \nabla \times \mathbf{B})] + \hat{r} \cdot\mathbf{S})\label{eq:poloidal_evolve}\\
    \frac{\partial A}{\partial t} &=& -r^2 \mathcal{L}^{-2}(\hat{r} \cdot \nabla\times[\nabla \times (\mathbf{V} \times \mathbf{B} - \eta \nabla \times \mathbf{B})] + \hat{r} \cdot \nabla \times \mathbf{S})\label{eq:toroidal_evolve},
\end{eqnarray}
where the longitudinal operator $\mathcal{L}$ is defined in spherical coordinates as
\begin{equation}
    \mathcal{L}^2 =
\left[
\frac{1}{\sin\theta}\frac{\partial}{\partial\theta}
\left( \sin\theta\,\frac{\partial}{\partial\theta} \right)
+ \frac{1}{\sin^2\theta}\frac{\partial^2}{\partial\phi^2}
\right],
\end{equation}
In our present formalism, we impose source term for poloidal field only. Therefore, $\hat{r} \cdot \nabla \times \mathbf{S} = 0.$ Following Equation~\ref{eq:poloidal_evolve}, the poloidal source term $S_C$ then becomes 
\begin{eqnarray}
    S_C &=&-r^2\mathcal{L}^{-2}(\hat{r} \cdot\mathbf{S})\\
        &=& -r^2\mathcal{L}^{-2} (B_{r, extrapol}).
\end{eqnarray}

Taking spherical harmonic transformations in latitude and longitude, this relation leads to the Equation~\ref{eq:poloidal_pot_compute}, the expression for the $S_C$.

\section{Incorporating helioseismic differential rotation data and meridional flow}\label{sec:mcflow}

\begin{figure}[!htbp]
   \centering
   \includegraphics[width=0.8\linewidth]{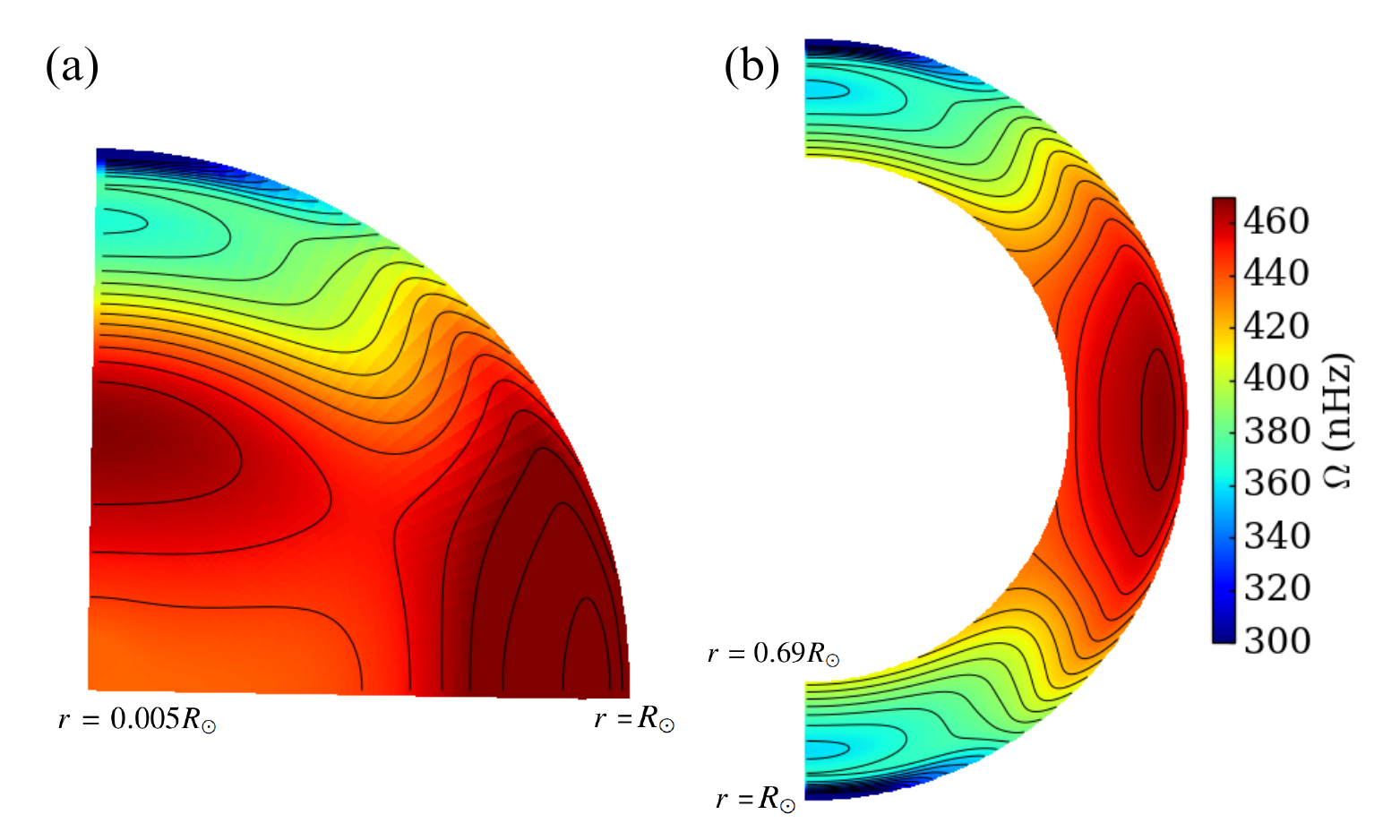}
   \caption{(a) Differential rotation profile, \(\Omega\), derived from helioseismology data by MDI on 1st May 1996. (b) The profile trimmed to the range \(r = 0.69R_\odot\) to \(R_\odot\), and extended over \(\theta \in [-90^\circ, 90^\circ]\) for use in our dynamo model.}
   \label{Fig:assimilated_dfr_mdi}%
\end{figure}

Here we describe how the velocity field $\mathbf{V}$ used in Equation~\ref{eq:induction_eq} has been incorporated in our 3D dynamo model.  The velocity field has two components: differential rotation and meridional circulation. It is defined as
\begin{equation}
    \mathbf{V} =  \tilde{\rho}(r)^{-1} \nabla \times [ \psi(r, \theta) \hat{\phi} ] + \lambda\Omega \hat{\phi},
\end{equation}
where $\tilde{\rho}(r)$ is the dimensionless density stratification, $\psi$ is the stream function describing the meridional flow, and $\Omega$ is the angular velocity representing the differential rotation. The profile of $\Omega$ is taken from helioseismic observations based on 72-day time series from MDI and HMI \citep{Antia_2022}, which we obtained through private communication from Prof. H M Antia. The dataset spans $r = 0.005R_\odot$ to $R_\odot$ in radial steps of $0.005R_\odot$, and covers latitudes from $0^\circ$ to $88^\circ$ with 2-degree spacing. Previous studies, such as \citet{Karak_2016_apj}, used a hybrid approach that combined helioseismic and analytical profiles to handle uncertainties at high latitudes. In contrast, we use the observed profile directly up to $88^\circ$ latitude, as the data provided are considered reliable within this range. For latitudes beyond $88^\circ$, we apply smooth extrapolation and extend the profile across the full latitudinal range $\theta \in [-90^\circ, 90^\circ]$ using interpolation with 256 grid points. Figures~\ref{Fig:assimilated_dfr_mdi}(a,b) show both the original and extended $\Omega$ profiles used in our simulations.
To include the meridional circulation, we use stream function $\psi$ following the formulation of \citep{HKC14}:
\begin{eqnarray}
    \psi r\sin\theta = \psi_0 (r -R_p) \sin\left[\frac{\pi(r-R_p)}{(R_\odot-R_p)}\right]
    (1 - e^{-\beta_1 r \theta^\epsilon})(1 - e^{-\beta_2 (\theta - \pi/2)}) e^{-((r - r_0)/\Gamma)^2},
\end{eqnarray}
where $\beta_1 = 1.0$, $\beta_2 = 1.3$, $\epsilon = 2.0000001$, $r_0 = 0.45R_\odot/3.5$, $\Gamma = 3.47 \times 10^{8}\,\mathrm{cm}$, and $R_p = 0.7R_\odot$. We set $\psi_0 = 38$ to achieve a peak surface meridional circulation speed of $19\,\mathrm{ms^{-1}}$. 

\section{Turbulent diffusivity} \label{sec:diffusivity}

We use a two-step profile for $\eta$, whose analytical form is given by \citep{HCM17}:
\begin{equation}
    \eta = \eta_c + \frac{\eta_{mid}}{2} \left[1 + \text{erf}\left(2\frac{r - r_{da}}{d_a}\right)\right] + \frac{\eta_{top}}{2} \left[1 + \text{erf}\left(2\frac{r - r_{db}}{d_b}\right)\right],
\end{equation}
where $\eta_c$, $\eta_{mid}$, and $\eta_{top}$ are the diffusivities at the bottom, mid, and top of the convection zone, respectively. We set $\eta_{top} = 3\times10^{12}\,\text{cm}^2\text{s}^{-1}$ to remain consistent with observations and established SFT models \citep{Jiang2014SSRv}. In addition, we choose $\eta_{mid} = 5\times10^{11}\,\text{cm}^2\text{s}^{-1}$, $\eta_c = 2\times10^{10}\,\text{cm}^2\text{s}^{-1}$, $r_{da} = 0.735R_\odot$, $r_{db} = 0.956R_\odot$, $d_a = 0.021R_\odot$, and $d_b = 0.05R_\odot$. Note that we use a high value of $\eta_{mid}$ to ensure that diffusion dominates over advection by the meridional flow within the convection zone \citep{yeates2008}. We ran a few cases with different diffusivities in the convection zone by varying $\eta_c$ and $\eta_{mid}$. An order of magnitude reduction in the diffusivity coefficients of the bottom and middle convection zones increases the peak field strength approximately to 6.5 kG at the bottom of the convection zone.  

\section{Constraining cross-equatorial diffusion of surface magnetic flux} \label{sec:cross-equtorial}

While comparing the toroidal flux estimated from our model with another way of constraining toroidal flux proposed by \citet{cameron2015} using Stokes' theorem, we find that our toroidal flux matches qualitatively with their prescription but it has some deviation, especially during solar maxima. We explain the reason below. \citet{cameron2015} demonstrated that one can estimate the generation of toroidal magnetic flux in the convection zone directly from the observed surface radial magnetic field and the differential rotation profile. This can be expressed as:

\begin{equation}
    \frac{d\Phi_N}{dt} = I_N, \label{eq:cameron}
\end{equation}
where $\Phi_N$ is the net toroidal flux generated in the northern hemisphere of the convection zone, and
\begin{equation}
    I_N = \int_{0}^{\pi/2} \langle V_{\phi} \rangle_\phi \langle B_{r} \rangle_\phi \, d\theta, \label{eq:IN}
\end{equation}
where $\langle V_{\phi} \rangle_\phi$ and $\langle B_{r} \rangle_\phi$ represent the azimuthally averaged surface angular velocity and radial magnetic field, respectively. The integral is evaluated over the azimuthally averaged surface of the northern hemisphere. A similar definition ($I_S$) applies to the southern hemisphere. 


The results are shown in Figure~\ref{fig:tor_flux}(a). The net toroidal flux generation in the northern and southern hemispheres, $\Phi_N$ and $\Phi_S$, is shown using green and blue solid lines, respectively. The corresponding surface integrals ($I_N$ (north), Equation~\ref{eq:IN}) from the right-hand side of Equation~\ref{eq:cameron}, integrated over time $\int{I_Ndt}$ and  $\int{I_Sdt}$ are evaluated using the assimilated surface magnetic field and differential rotation on both hemispheres and are plotted using dashed green and blue lines. The comparison shows that surface quantities alone can approximate the order of magnitude of the net toroidal flux, although the values do not match exactly.

\begin{figure}[!htbp]
   \centering
   \includegraphics[width=0.95\linewidth]{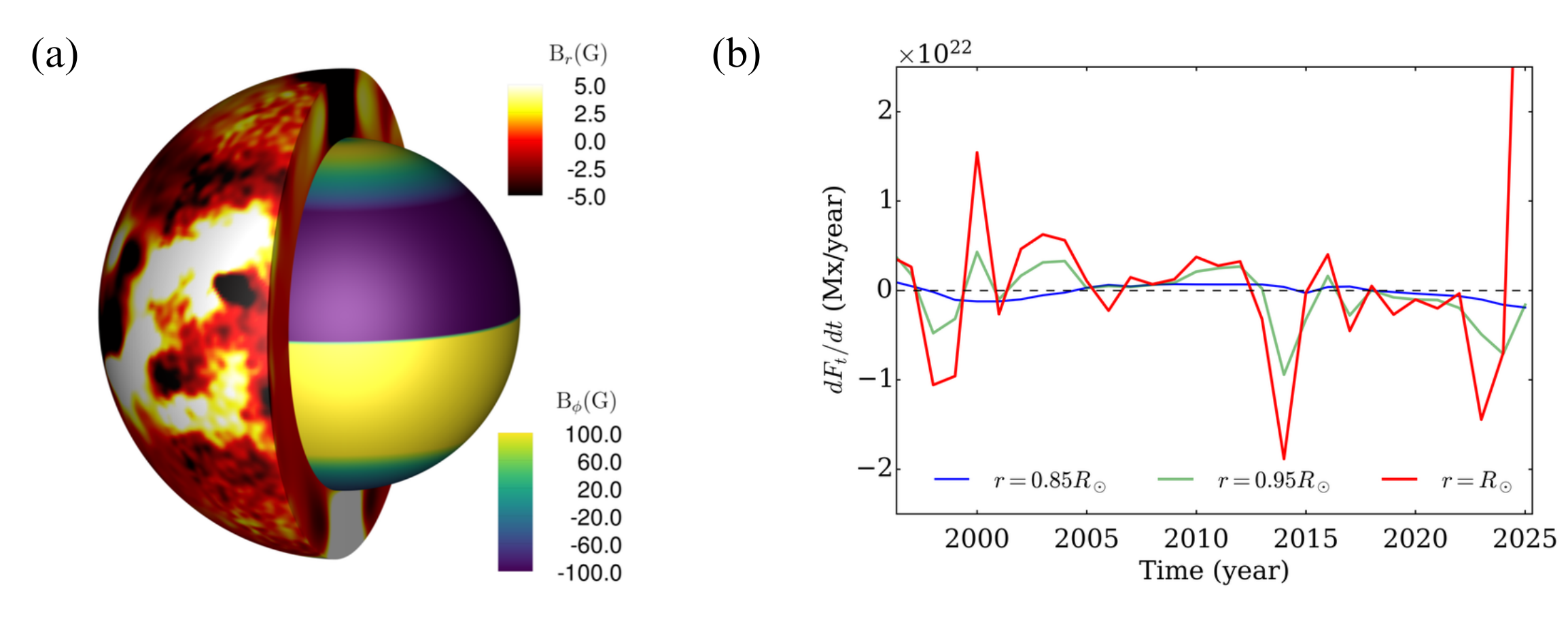}
   \caption{(a) Visualization of $B_r$ showing cross-equatorial transport at the surface. In contrast, the toroidal field $B_\phi$ near the tachocline does not exhibit any cross-equatorial transfer. (b) Time evolution of the net hemispheric magnetic flux transport rate, $\mathrm{d}F_t/\mathrm{d}t$, at different depths: $r = 0.85R_\odot$ (blue), $0.95R_\odot$ (green) and $R_\odot$ (red).}
   \label{Fig:cross_equatorial}%
\end{figure}

This discrepancy arises from the presence of substantial cross-equatorial magnetic flux transport. Figure~\ref{Fig:cross_equatorial}(a) illustrates the cross-equatorial transfer of $B_r$ at the solar surface, while panel (b) quantifies this transport at different depths. Following \citet{Cameron_2013A&A}, we define the flux imbalance at a given radius in the northern hemisphere as
\begin{equation}
    F_t = \frac{F_{NH} - F_{SH}}{2}, \label{eq:crosstransfer}
\end{equation}
where $F_{NH}$ and $F_{SH}$ are the total signed radial fluxes in the northern and southern hemispheres, respectively. The cross-equatorial transport is then computed as $dF_t/dt$, which is shown in Figure~\ref{Fig:cross_equatorial}(b) as a function of time. We see the cross-equatorial diffusion becomes stronger near surface and during solar maxima as shown by red solid line in Figure~\ref{Fig:cross_equatorial}(b). As we go deep inside the convection zone, cross-equatorial diffusive flux becomes less (see blue solid line). 


A critical examination might attribute the initial deviation observed in Figure~\ref{fig:tor_flux} due to the "SOHO summer vacation" data gap. However we are sure that the deviation in the all three cycles must be due to the enhanced diffusion of flux across equator, as enhanced cross-equatorial diffusion is observed during solar maxima in Figure~\ref{Fig:cross_equatorial}, aligning with the periods where the deviation from the theoretical estimate is most prominent.

\section{Effect of non-axisymmetric component of toroidal field}\label{sec:nas_comp}

\begin{figure}[!htbp]
   \centering
   \includegraphics[width=\linewidth]{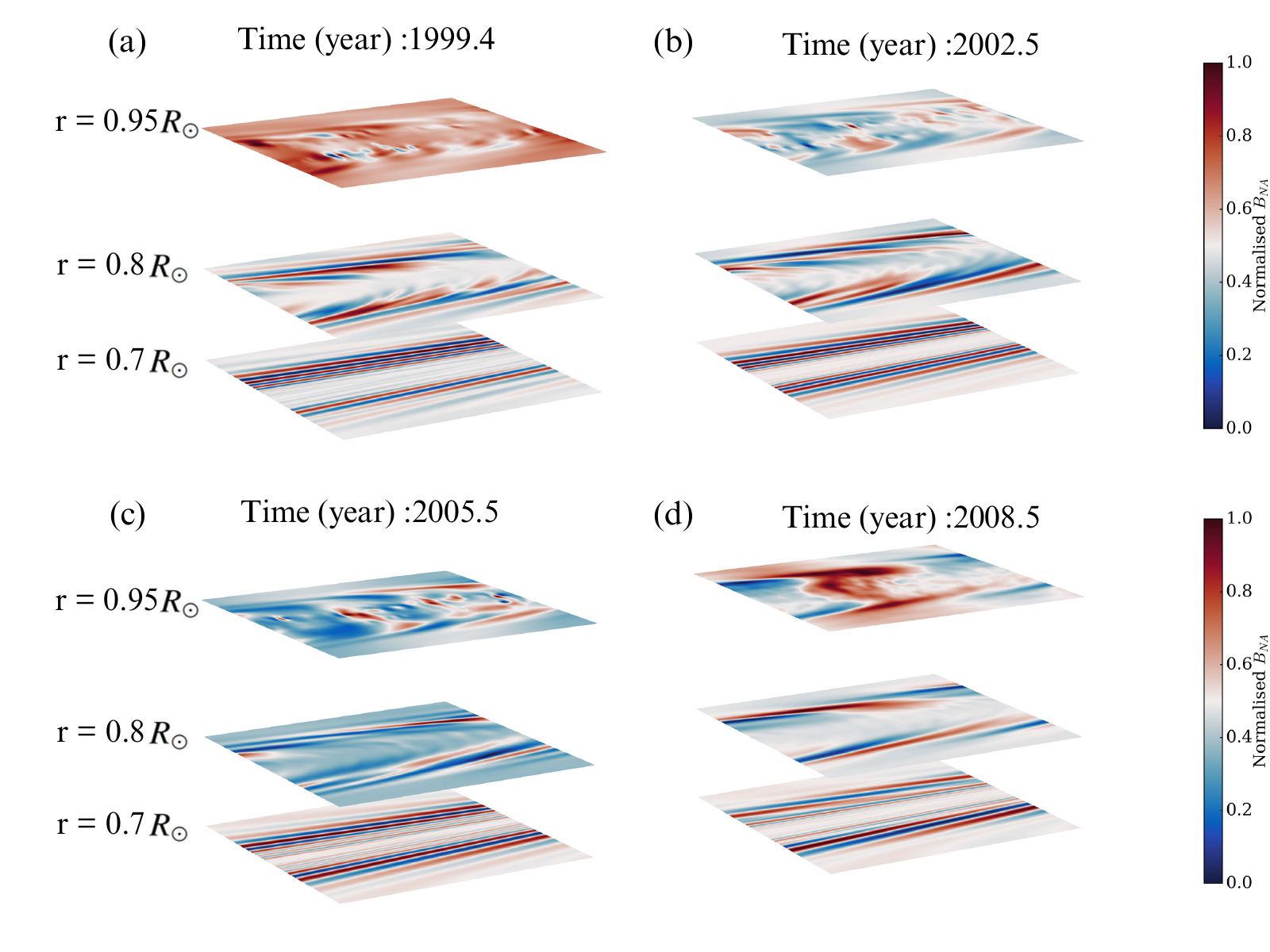}
   \caption{(a) Non-axisymmetric component at three depths, $r = 0.95R_\odot$, $0.8R_\odot$, and $0.7R_\odot$ from top to bottom at year 1999.4, corresponding to the rising phase of cycle 23. Other panel (b), (c) and (d) show the same thing as panel (a) but at maximum (2002.5), declining phase (2005.5), and minimum (2008.5) of Solar Cycle 23, respectively.}
   \label{Fig:nonaxisym_layers}%
\end{figure}

We also examine whether surface measurements are sufficient to estimate the toroidal field without applying azimuthal averaging. The motivation behind this is to account for the contribution from non-axisymmetric components. Figure~\ref{Fig:nonaxisym_layers} illustrates the non-axisymmetric component of the toroidal field at three depths ($r = 0.95R_\odot$, $0.8R_\odot$, and $0.7R_\odot$) and four different times (at years 1999.4, 2002.5, 2005.5, and 2008.5), spanning rising, maximum, declining, and minimum phases of Solar Cycle 23. These snapshots reveal that non-axisymmetric structures are predominantly concentrated near the solar surface and diminish progressively with depth. We also observe mixing of non-axisymmetric components. Since the Babcock–Leighton mechanism is implemented without a self-consistent flux emergence process, the deeper layers remain largely axisymmetric. In contrast, the upper layers near the surface exhibit stronger non-axisymmetric features due to the extrapolation of the observed surface magnetic field. The possibility of this is mentioned in ~\citet{Berdyugina_2006A&A}. This is also consistent with findings from mean-field models such as those by \citet{Belda_2017AA}, where the non-axisymmetric component was found to contribute only a few percent of the total azimuthal flux. Interestingly, the differences in non-axisymmetric structures between the years 1999.4 and 2005.5, the rising and declining phases of the cycle, suggest time-dependent complexity in subsurface magnetic organization. This point needs further investigation, as our current model does not fully capture the non-axisymmetric flux emergence.

\bibliographystyle{aasjournalv7}
\bibliography{ms, our_ref}


\end{document}